\documentclass[12pt]{article}
\usepackage{graphicx}
\usepackage{cite}
\newcommand{\ba}{\begin{array}}
\newcommand{\ea}{\end{array}}
\newcommand{\be}{\begin{equation}}
\newcommand{\ee}{\end{equation}}
\makeatletter 

\begin{document}
\baselineskip=24pt
\thispagestyle{empty}
\topskip=0.5cm
\begin{flushright}
\begin{tabular}{c}
\end{tabular}
\end{flushright}
\vspace{1cm}
\begin{center}

{\Large\bf $S_{3}$-flavour symmetry as realized in lepton flavour
  violating processes.}
\vspace{1cm}
\linebreak
A. Mondrag\'on, M. Mondrag\'on and E. Peinado\footnote{
mondra@fisica.unam.mx\\
\hspace*{0.5cm}myriam@fisica.unam.mx\\
\hspace*{0.5cm}eduardo@fisica.unam.mx} 


\vspace{0.2in}

{\sl Instituto de    F\'{\i}sica, Universidad Nacional Aut\'onoma
  de M\'exico,
    Apdo. Postal 20-364,  01000 M{\'e}xico D.F., \ M{\'e}xico.}
\end{center}

\vspace{0.7cm}

\begin{abstract}
A variety of lepton flavour violating effects related to the recent
discovery of neutrino oscillations and mixings is here systematically
discussed in terms of an $S_{3}$-flavour permutational symmetry. After a brief review of some relevant results on lepton masses and
mixings, that had been derived in the framework of a Minimal
$S_3$-Invariant Extension of the Standard Model, we derive
explicit analytical expressions for the matrices of the Yukawa
couplings and compute the branching ratios of some selected flavour
changing neutral current (FCNC) processes, as well as, the contribution of the
exchange of neutral flavour changing scalars to the anomaly of the
muon's magnetic moment as functions of the masses of the
charged leptons and the neutral Higgs bosons. We find that the
$S_3\times Z_2$ flavour symmetry and the strong mass hierarchy of the charged
leptons strongly suppress the FCNC processes in the leptonic sector
well below the present experimental upper bounds by many orders of
magnitude. The contribution of FCNC to the anomaly of the muon's
magnetic moment is small but non-negligible.


\end{abstract}
\begin{center}
PACS numbers: 11.30.Hv,14.60.Pq,14.60.St,14.80.Cp,12.15.Ff,12.15.Mm
\end{center}

\section{\label{int}Introduction}
Neutrino oscillation observations and experiments, made in the past
nine years, have allowed the determination of the differences of the
squared neutrino masses and the mixing angles in the leptonic
sector~\cite{jung,Mohapatra:2006gs,altmann,smy,ahmad,aharmim,fukuda,Ashie:2005ik,bemporad,Araki:2004mb,Maltoni:2004ei,schwetz,Gonzalez-Garcia:2007ib,chooz,eitel,eliot,Seljak,elgaroy,Lesgourgues}. The discovery that neutrinos have non-vanishing masses
and mix among themselves much like the quarks do, provides the first
conclusive evidence of new physics beyond the Standard Model. This
important discovery also brought out very forcefully the need of
extending the Standard Model to accommodate in the theory the new data
on neutrino physics in a coherent way, free of contradictions, and
without spoiling the Standard Model's many phenomenological successes.

In the Standard Model, the Higgs and Yukawa sectors, which are
responsible for the generation of the masses of quarks and charged
leptons, do not give mass to the neutrinos. Furthermore, the Yukawa
sector of the Standard Model already has too many parameters
whose values can only be determined from experiment. These two facts
point to the necessity and convenience of extending the Standard Model
in order to make a unified and
systematic treatment of the observed hierarchies of
masses and mixings of all fermions, as well as the presence or absence of CP violating
phases in the mixing matrices. At the same time, we would also like to
reduce drastically the number of free parameters in the theory. These
two seemingly contradictory demands can be met by means of a flavour
symmetry under which the families transform in a non-trivial fashion.

Recently, we argued that such a flavour
symmetry unbroken at the Fermi scale, is the permutational symmetry of
three objects $S_{3}$, and introduced a minimal $S_{3}$-invariant Extension of the
Standard Model~\cite{kubo1}. In this model, we imposed $S_{3}$ as a
fundamental symmetry in the matter sector. This assumption led us
necessarily to extend the concept of flavour and generations to the
Higgs sector. Hence, going to the irreducible representations of
$S_{3}$, we added to the Higgs $SU(2)_{L}$ doublet in the
$S_{3}$-singlet representation two more Higgs $SU(2)_{L}$
doublets, which can only belong to the two components of the
$S_{3}$-doublet representation, in this way, all the matter fields in
the Minimal $S_{3}$-invariant Extension of the Standard Model - Higgs,
quark and lepton fields, including the right handed neutrino fields-
belong to the three dimensional representation ${\bf 1}\oplus{\bf 2}$
of the permutational group $S_{3}$. The leptonic sector of the model
was further constrained by an Abelian $Z_{2}$ symmetry. We found that
the $S_3 \times Z_2$ symmetry predicts an almost maximal $\sin \theta_{23}$ and a very small
value for $\sin \theta_{13}$ and an
inverted mass hierarchy of the left handed neutrinos in good agreement
with experiment~\cite{kubo1,Kubo:2005sr}. More recently, we
reparametrized the mass matrices of the charged leptons and neutrinos,
previously derived in~\cite{kubo1}, in terms of their eigenvalues and
computed the neutrino mixing matrix, $V_{PMNS}$, and the neutrino
mixing angles and Majorana phases as functions of the masses of
charged leptons and neutrinos. The numerical values of the reactor, $\theta_{13}$, and atmosferic,
$\theta_{23}$, mixing angles are determined only by the masses of the
charged leptons in very good agreement with experiment. The solar
mixing angle, $\theta_{12}$, is almost insensitive to the values of
the masses of the charged leptons, but its experimental value allowed
us to fix the scale and origin of the neutrino mass spectrum. We found
that the theoretical neutrino mixing matrix $V_{PMNS}$ is nearly
tri-bimaximal in excellent agreement with the latest
experimental values~\cite{Felix:2006pn,Mondragon:2007af}.

The symmetry
$S_{3}$~\cite{Fritzsc1,pakvasa1,Fritzsc2,harari,Frere:1978ds,Fritzsc3,yamanaka,kaus,Fritzsch4,Harrison}
and the symmetry product
groups $S_{3}\times S_{3}$~\cite{Harrison,mondragon1,mondragon2,xing} and
$S_{3}\times S_{3}\times S_{3}$~\cite{hall,hall2} broken at the Fermi
scale, have been
considered by many authors to explain successfully the hierarchical
structure of quark masses and mixings in the Standard Model. Some other interesting models based on the $S_{3}$,
$S_{4}$, $A_{4}$ and $D_5$ flavour symmetry groups, unbroken at the Fermi
scale, have also been proposed~\cite{koide,ma,ma2,babu,chen,grimus-la,Hagedorn:2006ug,Hagedorn:2006ir}. Recent flavour symmetry models are
reviewed in
~\cite{Smirnov:2006qz,Altarelli:2004za,Mondragon:2006hi,Albright:2006cw},
see also the references therein.

In this paper, after a short, updated review of some relevant results
on lepton masses and mixings, we had previously derived, we will
discuss some other important flavour violating effects in the minimal
$S_3$-Invariant extension of the Standard 
Model. We will give exact explicit expressions for the matrices of the
Yukawa couplings in the leptonic sector expressed as functions of the masses of charged
leptons and neutral Higgs bosons. With the help of the Yukawa matrices
we will compute the branching ratios of some selected FCNC processes
and the contribution of the exchange of neutral flavour changing
scalars to the anomaly of the muon's magnetic moment. We find that the interplay of the
$S_3\times Z_2$ flavour symmetry and the strong mass hierarchy of
charged leptons strongly suppress the FCNC processes in the leptonic
sector well below the experimental upper bounds by many orders of
magnitude. The contribution to the anomaly, $a_{\mu}$, from FCNC is at
most $6\%$ of the discrepancy between the experimental value and the
Standard Model prediction for $a_{\mu}$, which is a small but not
negligible contribution.

\section{The Minimal $S_{3}$-invariant Extension of the Standard
  Model}
In the Standard Model analogous fermions in different generations have
identical couplings to all gauge bosons of the strong, weak
and electromagnetic interactions. Prior to the introduction of the
Higgs boson and mass terms, the Lagrangian is chiral and invariant
with respect to permutations of the left and right fermionic fields.

The six possible permutations of three objects $(f_{1},f_{2},f_{3})$
are elements of the permutational group $S_{3}$. This is the discrete,
non-Abelian group with the smallest number of elements. The
three-dimensional real representation is not an irreducible
representation of $S_{3}$. It can be decomposed into the direct
sum of a doublet $f_{D}$ and a singlet $f_{s}$, where
\be
\begin{array}{l}
f_{s}=\frac{1}{\sqrt{3}}(f_{1}+f_{2}+f_{3}),\\
\\
f_{D}^{T}=\left(\frac{1}{\sqrt{2}}(f_{1}-f_{2}),\frac{1}{\sqrt{6}}(f_{1}+f_{2}-2f_{3})\right).
\end{array}
\ee
The direct product of two doublets ${\bf p_{D}}^{T} =(p_{D1},p_{D2})$
and ${\bf q_{D}}^{T}=(q_{D1},q_{D2})$ may be decomposed into the direct
sum of two singlets ${\bf r_{s}}$ and ${\bf r_{s'}}$, and one doublet
${\bf r_{D}}^{T}$ where
\be
\begin{array}{lr}
{\bf r_{s}} = p_{D1} q_{D1} + p_{D2}q_{D2}, & {\bf r_{s'}} =
p_{D1}q_{D2} - p_{D2}q_{D1},
\end{array}
\ee
\be
{\bf r_{D}}^{T}= (r_{D1},r_{D2})=(p_{D1} q_{D2} + p_{D2}q_{D1},p_{D1} q_{D1} - p_{D2}q_{D2}).
\ee
The antisymmetric singlet ${\bf r_{s'}}$ is not invariant under $S_{3}$.

Since the Standard Model has only one Higgs $SU(2)_{L}$ doublet,
which can only be an $S_{3}$ singlet, it can only give mass to the
quark or charged lepton in the $S_{3}$ singlet representation, one in
each family, without breaking the $S_{3}$ symmetry.

Hence, in order to impose $S_{3}$ as a fundamental symmetry, unbroken
at the Fermi scale, we are led to extend the Higgs sector of the
theory. The quark, lepton and Higgs fields are
\be
\begin{array}{lr}
Q^T=(u_L,d_L)~,~ u_R~,~d_R~,~\\L^T=(\nu_L,e_L)~,~e_R~,~ 
\nu_R~\mbox{ and }~H,
\end{array}
\ee
in an obvious notation. All of these fields have three species, and
we assume that each one forms a reducible representation ${\bf 1}_S\oplus{\bf 2}$.
The doublets carry capital indices $I$ and $J$, which run from $1$ to $2$,
and the singlets are denoted by
$Q_3,~u_{3R},~d_{3R},~L_3,~e_{3R},~\nu_{3R}$ and $~H_S$. Note that the subscript $3$ denotes the
singlet representation and not the third generation.
The most general renormalizable Yukawa interactions of this model are given by
\be
{\cal L}_Y = {\cal L}_{Y_D}+{\cal L}_{Y_U}
+{\cal L}_{Y_E}+{\cal L}_{Y_\nu},
\ee
where
\be
\begin{array}{lll}
{\cal L}_{Y_D} &=&
- Y_1^d \overline{ Q}_I H_S d_{IR} - Y_3^d \overline{ Q}_3 H_S d_{3R}  \\
&  &   -Y^{d}_{2}[~ \overline{ Q}_{I} \kappa_{IJ} H_1  d_{JR}
+\overline{ Q}_{I} \eta_{IJ} H_2  d_{JR}~]\\
&  & -Y^d_{4} \overline{ Q}_3 H_I  d_{IR} - Y^d_{5} \overline{ Q}_I H_I d_{3R} 
+~\mbox{h.c.} ,
\label{lagd}
\end{array}
\ee
\be
\begin{array}{lll}
{\cal L}_{Y_U} &=&
-Y^u_1 \overline{ Q}_{I}(i \sigma_2) H_S^* u_{IR} 
-Y^u_3\overline{ Q}_3(i \sigma_2) H_S^* u_{3R} \\
&  &   -Y^{u}_{2}[~ \overline{ Q}_{I} \kappa_{IJ} (i \sigma_2)H_1^*  u_{JR}
+  \overline{ Q}_{I} \eta_{IJ}(i \sigma_2) H_2^*  u_{JR}~]\\
&  &
-Y^u_{4} \overline{ Q}_{3} (i \sigma_2)H_I^* u_{IR} 
-Y^u_{5}\overline{ Q}_I (i \sigma_2)H_I^*  u_{3R} +~\mbox{h.c.},
\label{lagu}
\end{array}
\ee
\be
\begin{array}{lll}
{\cal L}_{Y_E} &=& -Y^e_1\overline{ L}_I H_S e_{IR} 
-Y^e_3 \overline{ L}_3 H_S e_{3R} \\
&  &  - Y^{e}_{2}[~ \overline{ L}_{I}\kappa_{IJ}H_1  e_{JR}
+\overline{ L}_{I} \eta_{IJ} H_2  e_{JR}~]\\
 &  & -Y^e_{4}\overline{ L}_3 H_I e_{IR} 
-Y^e_{5} \overline{ L}_I H_I e_{3R} +~\mbox{h.c.},
\end{array}
\label{lage}
\ee
\be
\begin{array}{lcl}
{\cal L}_{Y_\nu} &=& -Y^{\nu}_1\overline{ L}_I (i \sigma_2)H_S^* \nu_{IR} 
-Y^\nu_3 \overline{ L}_3(i \sigma_2) H_S^* \nu_{3R} \\
&  &   -Y^{\nu}_{2}[~\overline{ L}_{I}\kappa_{IJ}(i \sigma_2)H_1^*  \nu_{JR}
+ \overline{ L}_{I} \eta_{IJ}(i \sigma_2) H_2^*  \nu_{JR}~]\\
 &  & -Y^\nu_{4}\overline{ L}_3(i \sigma_2) H_I^* \nu_{IR} 
-Y^\nu_{5} \overline{ L}_I (i \sigma_2)H_I^* \nu_{3R}+~\mbox{h.c.},
\label{lagnu}
\end{array}
\ee
and
\be
\kappa = \left( \begin{array}{cc}
0& 1\\ 1 & 0\\
\end{array}\right)~~\mbox{and}~~
\eta = \left( \begin{array}{cc}
1& 0\\ 0 & -1\\
\end{array}\right).
\label{kappa}
\ee Furthermore, we add to the Lagrangian the Majorana mass terms for
the right-handed neutrinos \be {\cal L}_{M} = -M_1 \nu_{IR}^T C
\nu_{IR} -M_3 \nu_{3R}^T C \nu_{3R}.
\label{majo}
\ee

Due to the presence of three Higgs fields, the Higgs potential
$V_H(H_S,H_D)$ is more complicated than that of the Standard
Model. A Higgs potential invariant under $S_{3}$ was first proposed by
Pakvasa and Sugawara~\cite{pakvasa1}, who assumed an additional
reflection symmetry $R:~H_{s}\to ~-H_{s}$. These authors found
that in addition to the $S_{3}$ symmetry, their Higgs potential has an
accidental permutational symmetry $S_{2}^{\prime}$:
$H_{1}\leftrightarrow H_{2}$. The accidental $S_{2}^{\prime}$ symmetry
is also present in our $V_H(H_S,H_D)$. The most general form of the
potential $V_H(H_S,H_D)$ was investigated in detail by Kubo, Okada and
Sakamaki~\cite{kubo-pot}, who discussed the potential of Pakvasa and Sugawara as a
special case. A preliminar study on conditions under which
the minimum of the Higgs potential is a global and stable one can be
found in ~\cite{EmmanuelCosta:2007zz}. In this communication, we will assume that the
vacuum respects the accidental $S_{2}^{\prime}$ symmetry of the Higgs potential
and therefore that
\be
\langle H_{1} \rangle = \langle H_{2} \rangle.
\ee

With these assumptions, the Yukawa interactions, eqs. (\ref{lagd})-(\ref{lagnu}) yield mass matrices,
for all fermions in the theory, of the general form~\cite{kubo1}
\be
{\bf M} = \left( \begin{array}{ccc}
\mu_{1}+\mu_{2} & \mu_{2} & \mu_{5} 
\\  \mu_{2} & \mu_{1}-\mu_{2} &\mu_{5}
  \\ \mu_{4} & \mu_{4}&  \mu_{3}
\end{array}\right).
\label{general-m}
\ee
The Majorana mass for the left handed neutrinos $\nu_{L}$ is generated
by the see-saw mechanism. The corresponding mass matrix is
given by
\be
{\bf M_{\nu}} = {\bf M_{\nu_D}}\tilde{{\bf M}}^{-1}({\bf M_{\nu_D}})^T,
\label{seesaw}
\ee
where $\tilde{{\bf M}}=\mbox{diag}(M_1,M_1,M_3)$.
\\
In principle, all entries in the mass matrices can be complex since
there is no restriction coming from the flavour symmetry $S_{3}$.
The mass matrices are diagonalized by bi-unitary transformations as
\be
\begin{array}{rcl}
U_{d(u,e)L}^{\dag}{\bf M}_{d(u,e)}U_{d(u,e)R} 
&=&\mbox{diag} (m_{d(u,e)}, m_{s(c,\mu)},m_{b(t,\tau)}),
\\ 
\\
U_{\nu}^{T}{\bf M_\nu}U_{\nu} &=&
\mbox{diag} (m_{\nu_1},m_{\nu_2},m_{\nu_3}).
\end{array}
\label{unu}
\ee
The entries in the diagonal matrices may be complex, so the physical
masses are their absolute values.

The mixing matrices are, by definition,
\be
\begin{array}{ll}
V_{CKM} = U_{uL}^{\dag} U_{dL},& V_{PMNS} = U_{eL}^{\dag} U_{\nu} K.
\label{ckm1}
\end{array}
\ee
where $K$ is the diagonal matrix of the Majorana phase factors.
\section{The mass matrices in the leptonic sector and $Z_{2}$
  symmetry}
A further reduction of the number of parameters in the leptonic sector
may be achieved by means of an Abelian $Z_{2}$ symmetry. A possible set
of charge assignments of $Z_{2}$, compatible with the experimental
data on masses and mixings in the leptonic sector is given in Table~\ref{table1}.

\begin{table}
\caption{\label{table1}$Z_2$ assignment in the leptonic sector.}
\begin{center}
\begin{tabular}{|c|c|}
\hline
 $-$ &  $+$
\\ \hline

$H_S, ~\nu_{3R}$ & $H_I, ~L_3, ~L_I, ~e_{3R},~ e_{IR},~\nu_{IR}$
\\ \hline
\end{tabular}
\end{center}
\end{table}

%
These $Z_2$ assignments forbid the following Yukawa couplings
\be
 Y^e_{1} = Y^e_{3}= Y^{\nu}_{1}= Y^{\nu}_{5}=0.
\label{zeros}
\ee
Therefore, the corresponding entries in the mass matrices vanish, {\it
  i.e.}, $\mu_{1}^{e}=\mu_{3}^{e}=0$ and $\mu_{1}^{\nu}=\mu_{5}^{\nu}=0$.
\begin{center}{\it The mass matrix of the charged leptons}\end{center}
The mass matrix of the charged leptons takes the form
\be
M_{e} = m_{\tau}\left( \begin{array}{ccc}
\tilde{\mu}_{2} & \tilde{\mu}_{2} & \tilde{\mu}_{5} 
\\  \tilde{\mu}_{2} &-\tilde{\mu}_{2} &\tilde{\mu}_{5}
  \\ \tilde{\mu}_{4} & \tilde{\mu}_{4}& 0
\end{array}\right).
\label{charged-leptons-m}
\ee
The unitary matrix $U_{eL}$ that enters in the definition of the
mixing matrix, $V_{PMNS}$, is calculated from
\be
U_{eL}^{\dag}M_{e}M_{e}^{\dag}U_{eL}=\mbox{diag}(m_{e}^{2},m_{\mu}^{2},m_{\tau}^{2}),
\ee
where $m_{e}$, $m_{\mu}$ and $m_{\tau}$ are the masses of the charged
leptons~\cite{Mondragon:2007af}.
The parameters $|\tilde{\mu}_{2}|$, $|\tilde{\mu}_{4}|$ and
$|\tilde{\mu}_{5}|$ may readily be expressed in terms of the charged
lepton masses~\cite{Felix:2006pn}. The resulting expression for $M_e$, written to order
$\left(m_{\mu}m_{e}/m_{\tau}^{2}\right)^{2}$ and
$x^{4}=\left(m_{e}/m_{\mu}\right)^4$ is
\be
M_{e}\approx m_{\tau} \left( 
\begin{array}{ccc}
\frac{1}{\sqrt{2}}\frac{\tilde{m}_{\mu}}{\sqrt{1+x^2}} & \frac{1}{\sqrt{2}}\frac{\tilde{m}_{\mu}}{\sqrt{1+x^2}} & \frac{1}{\sqrt{2}} \sqrt{\frac{1+x^2-\tilde{m}_{\mu}^2}{1+x^2}} \\ \\
 \frac{1}{\sqrt{2}}\frac{\tilde{m}_{\mu}}{\sqrt{1+x^2}} &-\frac{1}{\sqrt{2}}\frac{\tilde{m}_{\mu}}{\sqrt{1+x^2}}  & \frac{1}{\sqrt{2}} \sqrt{\frac{1+x^2-\tilde{m}_{\mu}^2}{1+x^2}} \\ \\
\frac{\tilde{m}_{e}(1+x^2)}{\sqrt{1+x^2-\tilde{m}_{\mu}^2}}e^{i\delta_{e}} & \frac{\tilde{m}_{e}(1+x^2)}{\sqrt{1+x^2-\tilde{m}_{\mu}^2}}e^{i\delta_{e}} & 0
\end{array}
\right).
\label{emass}
\ee
This approximation is numerically exact up to order $10^{-9}$ in units
of the $\tau$ mass. Notice that this matrix has no free parameters
other than the Dirac phase $\delta_e$.

The unitary matrix $U_{eL}$ that diagonalizes $M_{e}M_{e}^{\dagger}$ and
enters in the definition of the neutrino mixing matrix $V_{PMNS}$ may
be written as
\be
\ba{l}
U_{eL}= \left(\ba{ccc} 
1& 0 & 0 \\
0 & 1 & 0 \\
0 & 0 & e^{i\delta_{e}}
\ea\right) \left(
\ba{ccc}
O_{11}& -O_{12}& O_{13} \\
-O_{21}& O_{22}& O_{23} \\
-O_{31}& -O_{32}& O_{33} 
\ea
\right)~,
\ea
\label{unitary-leptons}
\ee
where the orthogonal matrix ${\bf O}_{eL}$ in the right hand side of
eq. (\ref{unitary-leptons}), written to the same order of magnitude as
$M_e$, is

\begin{eqnarray}
\hspace{-2.6cm}
{\bf O}_{eL}\approx
\left(
\ba{ccc}
\frac{1}{\sqrt{2}}x
\frac{(
1+2\tilde{m}_{\mu}^2+4x^2+\tilde{m}_{\mu}^4+2\tilde{m}_{e}^2
)}{\sqrt{1+\tilde{m}_{\mu}^2+5x^2-\tilde{m}_{\mu}^4-\tilde{m}_{\mu}^6+\tilde{m}_{e}^2+12x^4}}&
-\frac{1}{\sqrt{2}}\frac{(1-2\tilde{m}_{\mu}^2+\tilde{m}_{\mu}^4-2\tilde{m}_{e}^2)}{\sqrt{1-4\tilde{m}_{\mu}^2+x^2+6\tilde{m}_{\mu}^4-4\tilde{m}_{\mu}^6-5\tilde{m}_{e}^2}}
& \frac{1}{\sqrt{2}} \\ \\
-\frac{1}{\sqrt{2}}x
\frac{(
1+4x^2-\tilde{m}_{\mu}^4-2\tilde{m}_{e}^2
)}{\sqrt{1+\tilde{m}_{\mu}^2+5x^2-\tilde{m}_{\mu}^4-\tilde{m}_{\mu}^6+\tilde{m}_{e}^2+12x^4}}
&
\frac{1}{\sqrt{2}}\frac{(1-2\tilde{m}_{\mu}^2+\tilde{m}_{\mu}^4)}{\sqrt{1-4\tilde{m}_{\mu}^2+x^2+6\tilde{m}_{\mu}^4-4\tilde{m}_{\mu}^6-5\tilde{m}_{e}^2}}
& \frac{1}{\sqrt{2}} \\ \\
-\frac{\sqrt{1+2x^2-\tilde{m}_{\mu}^2-\tilde{m}_{e}^2}(1+\tilde{m}_{\mu}^2+x^2-2\tilde{m}_{e}^2)}{\sqrt{1+\tilde{m}_{\mu}^2+5x^2-\tilde{m}_{\mu}^4-\tilde{m}_{\mu}^6+\tilde{m}_{e}^2+12x^4}} & -x\frac{(1+x^2-\tilde{m}_{\mu}^2-2\tilde{m}_{e}^2)\sqrt{1+2x^2-\tilde{m}_{\mu}^2-\tilde{m}_{e}^2}}{\sqrt{1-4\tilde{m}_{\mu}^2+x^2+6\tilde{m}_{\mu}^4-4\tilde{m}_{\mu}^6-5\tilde{m}_{e}^2}} &\frac{\sqrt{1+x^2}\tilde{m}_{e}\tilde{m}_{\mu}}{\sqrt{1+x^2-\tilde{m}_{\mu}^2}}
\ea 
\right),\nonumber
\\
\label{unitary-leptons-2}
\end{eqnarray}

where, as before, $\tilde{m_{\mu}}=m_{\mu}/m_{\tau}$,
$\tilde{m_{e}}=m_{e}/m_{\tau}$ and $x=m_{e}/m_{\mu}$.
\begin{center}{\it The mass matrix of the neutrinos}\end{center}
According to the $Z_{2}$ selection rule eq. (\ref{zeros}), the mass
matrix of the Dirac neutrinos takes the form
\be
{\bf M_{\nu_D}} = \left( \begin{array}{ccc}
\mu^{\nu}_{2} & \mu^{\nu}_{2} & 0
\\  \mu^{\nu}_{2} & -\mu^{\nu}_{2} &0
  \\ \mu^{\nu}_{4} & \mu^{\nu}_{4}&  \mu^{\nu}_{3}
\end{array}\right).
\label{neutrinod-m}
\ee
Then, the mass matrix for the left-handed Majorana neutrinos, ${\bf
  M_{\nu}}$, obtained
from the see-saw mechanism, ${\bf M_{\nu}} = {\bf M_{\nu_D}}\tilde{{\bf M}}^{-1} 
({\bf M_{\nu_D}})^T$, is
\be
{\bf M_{\nu}} = 
\left( \begin{array}{ccc}
2 (\rho^{\nu}_{2})^2 & 0 & 
2 \rho^{\nu}_2 \rho^{\nu}_{4}
\\ 0 & 2 (\rho^{\nu}_{2})^2 & 0
  \\ 2 \rho^{\nu}_2 \rho^{\nu}_{4} & 0  &  
2 (\rho^{\nu}_{4})^2 +
(\rho^{\nu}_3)^2
\end{array}\right),
\label{m-nu}
\ee
where $\rho_2^\nu =(\mu^{\nu}_2)/M_{1}^{1/2}$,  $\rho_4^\nu
=(\mu^{\nu}_4)/M_{1}^{1/2}$ and $\rho_3^\nu
=(\mu^{\nu}_3)/M_{3}^{1/2}$; $M_{1}$ and $M_{3}$ are the masses of
the right handed neutrinos appearing in (\ref{majo}).

The non-Hermitian, complex, symmetric neutrino mass matrix $M_{\nu}$ may be brought
to a diagonal form by a unitary transformation, as
\be
U_{\nu}^{T}M_{\nu}U_{\nu}=\mbox{diag}\left(|m_{\nu_{1}}|e^{i\phi_{1}},|m_{\nu_{2}}|e^{i\phi_{2}},|m_{\nu_{3}}|e^{i\phi_{\nu}}\right),
\label{diagneutrino}
\ee
where $U_{\nu}$ is the matrix that diagonalizes the matrix
$M_{\nu}^{\dagger}M_{\nu}$.

As in the case of the charged leptons the matrices $M_{\nu}$ and
$U_\nu$, can be reparametrized in terms of the complex neutrino
masses. Then~\cite{Felix:2006pn,Mondragon:2007af}
\be
\hspace{-2.5cm}
M_{\nu} = 
\left( \begin{array}{ccc}
m_{\nu_{3}} & 0 & \sqrt{(m_{\nu_{3}}-m_{\nu_{1}})(m_{\nu_{2}}-m_{\nu_{3}})}e^{-i\delta_{\nu}}
\\ 
0 &m_{\nu_{3}}  & 0
\\
\sqrt{(m_{\nu_{3}}-m_{\nu_{1}})(m_{\nu_{2}}-m_{\nu_{3}})} e^{-i\delta_{\nu}} & 0  & (m_{\nu_{1}}+m_{\nu_{2}}-m_{\nu_{3}})e^{-2i\delta_{\nu}}
\end{array}\right)
\label{m-nu2}
\ee
and 
\be
U_{\nu}=\left(\ba{ccc} 
1& 0 & 0 \\
0 & 1 & 0 \\
0 & 0 & e^{i\delta_{\nu}} 
\ea\right)\left(
\begin{array}{ccc}
\cos \eta & \sin \eta & 0 \\
0 & 0 & 1 \\
-\sin \eta  & \cos \eta &0
\end{array}
\right),
\label{ununew}
\ee
where
\be
\ba{lr}
\sin^2\eta=\frac{m_{\nu_{3}}-m_{\nu_{1}}}{m_{\nu_{2}}-m_{\nu_{1}}},
&
\cos^2\eta=\frac{m_{\nu_{2}}-m_{\nu_{3}}}{m_{\nu_{2}}-m_{\nu_{1}}}.
\ea
\label{cosysin}
\ee
The unitarity of $U_{\nu}$ constrains $\sin \eta$ to be real and thus 
$|\sin \eta|\leq 1$, this condition fixes the phases $\phi_{1}$ and
$\phi_{2}$ as
\be
|m_{\nu_{1}}|\sin \phi_{1}=|m_{\nu_{2}}|\sin \phi_{2}=|m_{\nu_{3}}|\sin \phi_{\nu}.
\label{phase-condition}
\ee
The only free parameters in these matrices, are the phase $\phi_{\nu}$, implicit in $m_{\nu_{1}}$,
$m_{\nu_{2}}$ and $m_{\nu_{3}}$, and the Dirac phase $\delta_{\nu}$.
\bigskip
\begin{center}
{\it The neutrino mixing matrix}
\end{center}
The neutrino mixing matrix $V_{PMNS}$, is the product
$U_{eL}^{\dagger}U_{\nu}K$, where $K$ is the diagonal matrix of the
Majorana phase factors, defined by
\be
diag(m_{\nu_{1}},m_{\nu_{2}},m_{\nu_{3}})=K^{\dagger}diag(|m_{\nu_{1}}|,|m_{\nu_{2}}|,|m_{\nu_{3}}|)K^{\dagger}.
\ee
Except for an overall phase factor $e^{i\phi_{1}}$, which can be
ignored, $K$ is 
\be
K=diag(1,e^{i\alpha},e^{i\beta}),
\ee
where $\alpha=1/2(\phi_{1}-\phi_{2})$ and
$\beta=1/2(\phi_{1}-\phi_{\nu})$ are the Majorana phases. 

Therefore, the theoretical mixing matrix $V_{PMNS}^{th}$, is given by
\be
\hspace{-2.5cm}
V_{PMNS}^{th}=
\left(
\ba{ccc}
O_{11}\cos \eta + O_{31}\sin \eta e^{i\delta} & O_{11}\sin
\eta-O_{31} \cos \eta e^{i\delta} & -O_{21}  \\ \\
-O_{12}\cos \eta + O_{32}\sin \eta e^{i\delta} & -O_{12}\sin
\eta-O_{32}\cos \eta e^{i\delta} & O_{22} \\ \\
O_{13}\cos \eta - O_{33}\sin \eta e^{i\delta} & O_{13}\sin
\eta+O_{33}\cos \eta e^{i\delta} & O_{23} 
\ea
\right)
 \times K,
\label{vpmns2}
\ee
where $\cos \eta$ and $\sin \eta$ are given in eq (\ref{cosysin}),
$O_{ij}$ are given in eq (\ref{unitary-leptons}) and
(\ref{unitary-leptons-2}), and $\delta=\delta_{\nu}-\delta_{e}$. 

To find the relation of our results
with the neutrino mixing angles we make use of the equality of the
absolute values of the elements of $V_{PMNS}^{th}$ and
$V_{PMNS}^{PDG}$~\cite{PDG}, that is
\be
|V_{PMNS}^{th}|=|V_{PMNS}^{PDG}|.
\ee
This relation allows us to derive expressions for the mixing angles
in terms of the charged lepton and neutrino masses.

The magnitudes of the reactor and atmospheric mixing angles,
$\theta_{13}$ and $\theta_{23}$, are determined by the masses of the
charged leptons only. Keeping only terms of order $(m_{e}^2/m_{\mu}^2)$ and
$(m_{\mu}/m_{\tau})^4$, we get
\be
\ba{lr}
\sin \theta_{13}\approx \frac{1}{\sqrt{2}}x
\frac{(
1+4x^2-\tilde{m}_{\mu}^4)}{\sqrt{1+\tilde{m}_{\mu}^2+5x^2-\tilde{m}_{\mu}^4}}
, &
\sin \theta_{23}\approx  \frac{1}{\sqrt{2}}\frac{1+\frac{1}{4}x^2-2\tilde{m}_{\mu}^2+\tilde{m}_{\mu}^4}{\sqrt{1-4\tilde{m}_{\mu}^2+x^2+6\tilde{m}_{\mu}^4}}.
\ea
\ee
The magnitude of the solar angle depends on charged lepton and
neutrino masses, as well as, the Dirac and Majorana phases
\be
\hspace{-2.0cm}
 |\tan \theta_{12}|^2= \frac{\displaystyle{m_{\nu_{2}}-m_{\nu_{3}}}}{
\displaystyle{m_{\nu_{3}}-m_{\nu_{1}}}}\left(\frac{1-2\frac{O_{11}}{O_{31}}\cos \delta\sqrt{\frac{\displaystyle{m_{\nu_{3}}-m_{\nu_{1}}}}{
\displaystyle{m_{\nu_{2}}-m_{\nu_{3}}}}}+\left(\frac{O_{11}}{O_{31}}\right)^2\frac{\displaystyle{m_{\nu_{3}}-m_{\nu_{1}}}}{
\displaystyle{m_{\nu_{2}}-m_{\nu_{3}}}}}{1+2\frac{O_{11}}{O_{31}}\cos \delta\sqrt{\frac{\displaystyle{m_{\nu_{2}}-m_{\nu_{3}}}}{
\displaystyle{m_{\nu_{3}}-m_{\nu_{1}}}}}+\left(\frac{O_{11}}{O_{31}}\right)^2\frac{\displaystyle{m_{\nu_{2}}-m_{\nu_{3}}}}{
\displaystyle{m_{\nu_{3}}-m_{\nu_{1}}}}
}\right)
.
\label{tan2}
\ee

The dependence of $\tan \theta_{12}$ on the Dirac phase $\delta$, see
(\ref{tan2}), is very weak, since $O_{31}\sim 1$ but $O_{11}\sim
1/\sqrt{2}(m_e/m_\mu)$. Hence, we may neglect it when comparing
(\ref{tan2}) with the data on neutrino mixings.

The dependence of $\tan \theta_{12}$ on the phase $\phi_{\nu}$ and the
physical masses of the neutrinos enters through the ratio of the
neutrino mass differences, it can be made explicit with the help of
the unitarity constraint on $U_{\nu}$, 
eq. (\ref{phase-condition}),
\be
\frac{\displaystyle{m_{\nu_{2}}-m_{\nu_{3}}}}{
\displaystyle{m_{\nu_{3}}-m_{\nu_{1}}}}=
\frac{(|m_{\nu_{2}}|^2-|m_{\nu_{3}}|^{2}\sin^{2}\phi_{\nu})^{1/2}-|m_{\nu_{3}}|
  |\cos
    \phi_{\nu}|}
{(|m_{\nu_{1}}|^{2}-|m_{\nu_{3}}|^{2}\sin^{2}\phi_{\nu})^{1/2}+|m_{\nu_{3}}|
  |\cos
    \phi_{\nu}|}.
\label{tansq}
\ee
Similarly, the Majorana phases are given by
\be
\ba{l}
\sin 2\alpha=\sin(\phi_{1}-\phi_{2})=
\frac{|m_{\nu_{3}}|\sin\phi_{\nu}}{|m_{\nu_{1}}||m_{\nu_{2}}|}\times
\\
\left(\sqrt{|m_{\nu_{2}}|^2-|m_{\nu_{3}}|^{2}\sin^{2}\phi_{\nu}}+\sqrt{|m_{\nu_{1}}|^{2}-|m_{\nu_{3}}|^{2}\sin^{2}\phi_{\nu}}\right),
\ea
\ee
\be
\ba{l}
\sin 2\beta=\sin(\phi_{1}-\phi_{\nu})=
\\
 \frac{\sin\phi_{\nu}}{|m_{\nu_{1}}|}\left(|m_{\nu_{3}}|\sqrt{1-\sin^{2}\phi_{\nu}}+\sqrt{|m_{\nu_{1}}|^{2}-|m_{\nu_{3}}|^{2}\sin^{2}\phi_{\nu}}\right).
\ea
\ee
A more complete and detailed discussion of the Majorana phases in the
neutrino mixing matrix $V_{PMNS}$ obtained in our model is given by 
J. Kubo~\cite{kubo-u}.

\section{Neutrino masses and mixings}

In the present model, $\sin^{2} \theta_{13}$ and $\sin^{2} \theta_{23}$ are
determined by the masses of the charged leptons in very good
agreement with the experimental values~\cite{Maltoni:2004ei,schwetz,fogli1},
\be
\begin{array}{ll}
(\sin^{2}\theta_{13})^{th}=1.1\times 10^{-5}, &(\sin^2
  \theta_{13})^{exp} \leq 0.046, 
\end{array}
\ee
and
\be
\begin{array}{ll}
(\sin^{2}\theta_{23})^{th}=0.5, &(\sin^2
  \theta_{23})^{exp}=0.5^{+0.06}_{-0.05}.
\end{array}
\ee
In this model, the experimental restriction $|\Delta
m^2_{12}|<|\Delta m^2_{13}|$ implies an inverted neutrino mass
spectrum, $|m_{\nu_{3}}|<|m_{\nu_{1}}|<|m_{\nu_{2}}|$~\cite{kubo1}.

As can be seen from eqs. (\ref{tan2}) and (\ref{tansq}), the solar mixing angle is
sensitive to the neutrino mass differences and the phase $\phi_{\nu}$
but is only very weakly sensitive to the charged lepton
masses. If we neglect the small terms proportional to $O_{11}$ and $O_{11}^2$
in (\ref{tan2}), we get
\be
\begin{array}{l}
\tan^2 \theta_{12} =
\frac{(\Delta m_{12}^2+\Delta m_{13}^2+|m_{\nu_{3}}|^{2}\cos^{2}\phi_{\nu})^{1/2}-|m_{\nu_{3}}|
  |\cos
    \phi_{\nu}|}
{(\Delta m_{13}^2+|m_{\nu_{3}}|^{2}\cos^{2}\phi_{\nu})^{1/2}+|m_{\nu_{3}}|
  |\cos
    \phi_{\nu}|}.
\end{array}
\label{tansq2}
\ee

From this expression, we may readily derive expressions for the
neutrino masses in terms of $\tan \theta_{12}$ and $\phi_\nu$ and the
differences of the squared masses of the neutrinos masses
\be
|m_{\nu_{3}}|=\frac{\sqrt{\Delta m_{13}^2}}{2\cos \phi_{\nu} \tan
  \theta_{12}}\frac{1-\tan^4 \theta_{12}+r^2}{\sqrt{1+\tan^2 \theta_{12}} \sqrt{1+\tan^2 \theta_{12}+r^2}},
\label{masa3}
\ee
and 
\be
\ba{lr}
|m_{\nu_{1}}|=\sqrt{|m_{\nu_{3}}|^2+\Delta m_{13}^2},\hspace{1cm} & |m_{\nu_{2}}|=\sqrt{|m_{\nu_{3}}|^2+\Delta m_{13}^2(1+r^2)} 
\ea
\ee
where $r^2=\Delta m_{12}^2/\Delta m_{13}^2\approx 3\times 10^{-2}$. 

As $r^2<<1$, the sum of the neutrino masses is
\begin{eqnarray}
\hspace{-2.5cm}
\sum_{i=1}^{3} |m_{\nu_{i}}|\approx\frac{\sqrt{\Delta m_{13}^2}}{2\cos \phi_{\nu}
  \tan \theta_{12}}\left(1+2\sqrt{1+2\tan^2 \theta_{12}\left(2 \cos^2 \phi_\nu-1\right)+\tan^4 \theta_{12}}-\tan^2 \theta_{12}\right). \nonumber \\ 
\end{eqnarray}
The most restrictive cosmological upper bound for this sum is~\cite{Seljak}
\be
\sum |m_{\nu}|\leq 0.17 eV.
\ee
From this upper bound and the experimentally determined values of
$\tan \theta_{12}$ and $\Delta m_{ij}^{2}$, we may derive a lower
bound for $\cos \phi_{\nu}$
\be
\cos \phi_{\nu}\geq 0.55
\ee
or $0\leq \phi_\nu \leq 57^{\circ}$. The neutrino masses $|m_{\nu_i}|$
assume their minimal values when $\cos \phi_\nu=1$. When $\cos
\phi_\nu$ takes values in the range $0.55 \leq \cos \phi \leq 1$, the
neutrino masses change very slowly with $\cos \phi_\nu$. In the
absence of experimental information we will assume that $\phi_\nu$ vanishes.
Hence, setting $\phi_\nu=0$ in our formula, we find
\be
\ba{lcr}
|m_{\nu_{2}}|\approx0.056eV& |m_{\nu_{1}}|\approx 0.055eV&
|m_{\nu_{3}}|\approx 0.022eV,
\ea
\ee
where we used the values $\Delta m^{2}_{13}=2.6 \times 10^{-3}eV^{2}$,
$\Delta m^{2}_{21}= 7.9 \times 10^{-5}eV^{2}$
and $\tan \theta_{12}=0.667$, taken
from ~\cite{Gonzalez-Garcia:2007ib}.
\section{Flavour Changing Neutral Currents (FCNC)}

Models with more than one Higgs $SU(2)$ doublet have tree level
flavour changing neutral currents. In the Minimal
$S_{3}$-invariant Extension of the Standard Model, considered here,
there is one Higgs $SU(2)$ doublet per generation coupling to all fermions. The
flavour changing Yukawa couplings may be written in a flavour labelled,
symmetry adapted weak basis as
\be
\ba{lcl}
\hspace{-8pt}{\cal L}^{\rm FCNC}_{Y} =
\left(\overline{E}_{aL} Y_{a b}^{ES} E_{bR}+
\overline{U}_{aL} Y_{a b}^{US} U_{bR}
+\overline{D}_{aL} Y_{a b}^{DS} D_{bR}\right)H_S^0 \\ \\
~+\left(\overline{E}_{aL} Y_{a b}^{E1} E_{bR}+
\overline{U}_{aL} Y_{a b}^{U1} U_{bR}
+\overline{D}_{aL} Y_{a b}^{D1} D_{bR}\right)H_1^0+ \\ \\
\left(\overline{E}_{aL} Y_{a b}^{E2} E_{bR}+
\overline{U}_{aL} Y_{a b}^{U2} U_{bR}
+\overline{D}_{aL} Y_{a b}^{D2} D_{bR}\right)H_2^0+\mbox{h.c.}
\ea
\label{fcnf-lept}
\ee
where the entries in the column matrices $E's$, $U's$ and $D's$ are
the left and right fermion fields and $Y_{ab}^{(e,u,d)s}$,
$Y_{ab}^{(e,u,d)1,2}$ are $3\times 3$ matrices of the Yukawa couplings
of the fermion fields to the neutral Higgs fields $H_{s}^0$ and
$H_{I}^0$ in the the $S_3$-singlet and doublet representations, respectively.

In this basis, the Yukawa couplings of the Higgs fields to each
family of fermions may be written in terms of matrices
${\cal{M}}_{Y}^{(e,u,d)}$, which give rise to the corresponding mass
matrices $M^{(e,u,d)}$ when the gauge symmetry
is spontaneously broken. From this relation we may calculate the
flavour changing Yukawa couplings in terms of the fermion masses and
the vacuum expectation values of the neutral Higgs fields. For
example, the matrix ${\cal{M}}_{Y}^e$ is written in terms of the
matrices of the Yukawa couplings of the charged leptons as
\be
{\cal{M}}_{Y}^e=Y_{w}^{E1} H^0_{1}+Y_{w}^{E2} H^0_{2},
\ee
in this expression, the index $w$ means that the Yukawa matrices are
defined in the weak basis. The Yukawa couplings of immediate physical interest in the computation
of the flavour changing neutral currents are those defined in the mass
basis, according to
$\tilde{Y}_{m}^{EI}=U_{eL}^{\dagger}Y_{w}^{EI}U_{eR}$, where  $U_{eL}$ and
$U_{eR}$ are the matrices that diagonalize the charged lepton mass
matrix defined in eqs. (\ref{unu}) and (\ref{unitary-leptons}). We
obtain~\cite{{Mondragon:2007af}}
\be
\tilde{Y}_{m}^{E1}\approx \frac{m_{\tau}}{v_{1}}\left(
\ba{ccc}
2\tilde{m}_{e} & -\frac{1}{2}\tilde{m}_{e} & \frac{1}{2} x \\
\\
-\tilde{m}_{\mu} & \frac{1}{2}\tilde{m}_{\mu} & -\frac{1}{2} \\
\\
\frac{1}{2} \tilde{m}_{\mu} x^2 & -\frac{1}{2}\tilde{m}_{\mu} & \frac{1}{2}
\ea
\right)_{m},
\label{y1m}
\ee
and
\be
\tilde{Y}_m^{E2}\approx \frac{m_{\tau}}{v_{2}}\left(
\ba{ccc}
-\tilde{m}_{e} & \frac{1}{2}\tilde{m}_{e} & -\frac{1}{2} x \\
\\
\tilde{m}_{\mu} & \frac{1}{2}\tilde{m}_{\mu} & \frac{1}{2} \\
\\
-\frac{1}{2} \tilde{m}_{\mu} x^2 & \frac{1}{2}\tilde{m}_{\mu} & \frac{1}{2}
\ea
\right)_{m},
\label{y2m}
\ee
where $\tilde{m}_{\mu}=5.94\times 10^{-2}$, $\tilde{m}_{e}=2.876 \times
10^{-4}$ and $x=m_{e}/m_{\mu}=4.84 \times 10^{-3}$.
All the non-diagonal elements are responsible for tree-level FCNC
processes. The actual values of the Yukawa couplings in
eqs. (\ref{y1m}) and (\ref{y2m}) still depend on the VEV's of the
Higgs fields $v_{1}$ and $v_{2}$, and, hence, on the Higgs
potential. If the $S_{2}^{\prime}$ symmetry in the Higgs sector is
preserved~\cite{pakvasa1}, $\langle H_{1}^{0}
\rangle = \langle H_{2}^{0} \rangle= v $. In order
to make an order of magnitude estimate of the coefficient in the Yukawa
matrices, $m_\tau/v$, we may further assume that the VEV's
for all the Higgs fields are comparable, that is, $\tan \beta=\langle
H_{s}^{0} \rangle/\langle H_{1}^{0}\rangle=1$, and $\langle H_{s}^{0} \rangle=\langle H_{1}^{0}\rangle = \langle H_{2}^{0}
\rangle=\frac{\sqrt{2}}{\sqrt{3}}\frac{M_W}{g_2}$, then,  $m_\tau/v=\sqrt{3}/\sqrt{2}g_2m_\tau/M_W$
and we may estimate the numerical values of the Yukawa couplings from
the numerical 
\noindent values of the lepton masses. For instance, the amplitude of the
flavour violating process $\tau^-\to \mu^-e^+e^-$, is
proportional to $\tilde{Y}_{\tau \mu}^{E}\tilde{Y}_{e e}^{E}$~\cite{Sher:1991km}. Then,
the leptonic branching ratio,
\be
Br(\tau \to \mu e^+ e^-)=\frac{\Gamma(\tau \to \mu e^+
  e^-)}{\Gamma(\tau \to e \nu \bar{\nu})+\Gamma(\tau \to \mu \nu \bar{\nu})}
\ee
and 
\be
\Gamma(\tau \to \mu e^+
  e^-)\approx \frac{m_{\tau}^5}{3\times 2^{10} \pi^3}\frac{\left(Y^{1,2}_{ \tau
      \mu}Y^{1,2}_{ e e}\right)^2}{M_{H_{1,2}}^4}
\ee
which is the dominant term, and the well known expressions for
$\Gamma(\tau \to e \nu \bar{\nu})$ and $\Gamma(\tau
\to \mu \nu \bar{\nu})$~\cite{PDG}, give
\be
Br(\tau \to \mu e^+
e^-)\approx\frac{9}{4}\left(\frac{m_{e}m_{\mu}}{m_{\tau}^2}\right)^2
\left(\frac{m_{\tau}}{M_{H_{1,2}}}\right)^4,
\ee
taking for $M_{H_{1,2}}\sim 120~GeV$, we obtain $$Br(\tau
\to \mu e^+ e^-)\approx 3.15 \times 10^{-17}$$ well below the
experimental upper bound for this process, which is $2.7 \times
10^{-7}$~\cite{aubert}. 
\begin{table}
\caption{\label{table2}Leptonic FCNC processes, calculated with $M_{H_{1,2}}\sim 120~GeV$.}
\begin{center}
\begin{tabular}{|l|l|l|l|}
\hline
FCNC processes & Theoretical BR &  Experimental  & References \\
& & upper bound BR &
\\ \hline
$\tau \to  3\mu$ & $8.43
\times 10^{-14}$& $ 2 \times 10^{-7}$ & B. Aubert {\it et. al.} ~\cite{aubert}
 
\\ \hline
$\tau \to  \mu e^+ e^-$ & $3.15 \times 10^{-17}$& $2.7 \times 10^{-7} $ &B. Aubert {\it et. al.} ~\cite{aubert}
 
\\ \hline

$\tau \to \mu \gamma$ &  $9.24 \times 10^{-15}$ & $ 6.8 \times 10^{-8}$& B. Aubert {\it et. al.} ~\cite{aubert2} 
\\ \hline
$\tau \to e \gamma$ & $5.22\times 10^{-16}$ & $ 1.1 \times 10^{-11}$ &  B. Aubert {\it et. al.} ~\cite{aubert3}  
\\ \hline 
$\mu \to  3e$ &  $2.53 \times 10^{-16}$ & $  1 \times 10^{-12}$ &
U. Bellgardt {\it et al.} ~\cite{bellgardt}  
\\ \hline
$\mu \to e \gamma$ &  $2.42 \times 10^{-20}$ & $ 1.2 \times 10^{-11}$& M.~L.~Brooks {\it et al.} ~\cite{Brooks:1999pu}
\\\hline
\end{tabular}
\end{center}
\end{table}
Similar computations give the following estimates
\be
Br(\tau \to e
\gamma)\approx
\frac{3\alpha}{8\pi}\left(\frac{m_\mu}{M_H}\right)^4,
\label{tauegamma}
\ee
\be
Br(\tau \to \mu
\gamma)\approx\frac{3\alpha}{128\pi}\left(\frac{m_{\mu}}{m_{\tau}}\right)^2\left(\frac{m_\tau}{M_H}\right)^4,
\ee
\be
Br(\tau \to 3\mu)\approx\frac{9}{64}\left(\frac{m_\mu}{M_H}\right)^4,
\ee
\be
Br(\mu \to 3e)\approx 18 \left(\frac{m_e
    m_\mu}{m_\tau^2}\right)^2\left(\frac{m_\tau}{M_H}\right)^4,
\label{mu3e}
\ee
and 
\be
Br(\mu \to e
\gamma)\approx
\frac{27\alpha}{64\pi}\left(\frac{m_e}{m_\mu}\right)^4\left(\frac{m_\tau}{M_H}\right)^4.
\label{muegamma}
\ee
If we do not assume $v_{s}=v_{1}=v_{2}$, but keep $v_{s}/v_{1}=\tan
\beta$ unspecified, the expressions (\ref{tauegamma}-\ref{muegamma}) must
be multiplied by a factor $(2+\tan^2 \beta)^2/9$. 

We see that FCNC processes in the leptonic sector are strongly
suppressed by the small values of the mass ratios
$m_e/m_\tau$, $m_\mu/m_\tau$ and
$m_\tau/M_H$. The numerical estimates of the branching
ratios and the corresponding experimental upper bounds are shown in
Table~\ref{table2}. It may be seen that, in all cases considered, the numerical
values for the branching ratios of the FCNC in the leptonic sector are
well below the corresponding experimental upper bounds. 
The matrices of the quark Yukawa couplings may be computed in a
similar way. Numerical values for the Yukawa couplings for u and
d-type quarks are given in our previous paper ~\cite{kubo1}. There, it was
found that, due to the strong hierarchy in the quark masses and the
corresponding small or very small
 mass ratios, the numerical values of
all the Yukawa couplings in the quark sector are small or very
small. Kubo, Okada and Sakamaki~\cite{kubo-pot} have investigated the
breaking of the gauge symmetry in the case of the most general Higgs
potential invariant under $S_3$. They found that, by breaking the
$S_3$ symmetry very softly at very high energies it is possible to
maintain the consistency and predictions of the present
$S_{3}$-invariant Extension of the Standard Model while
simultaneously satisfying the experimental constraints for FCNC
processes, that is, it is possible that all physical Higgs bosons,
except one neutral one, could become sufficiently heavy
($M_H\sim 10~TeV$) to suppress all the flavour changing neutral
current processes in the quark sector of the theory without having a
problem with triviality.
\section{Muon anomalous magnetic moment}
\begin{center}
\begin{figure}[h]
\begin{center}
\includegraphics[width=6.5cm]{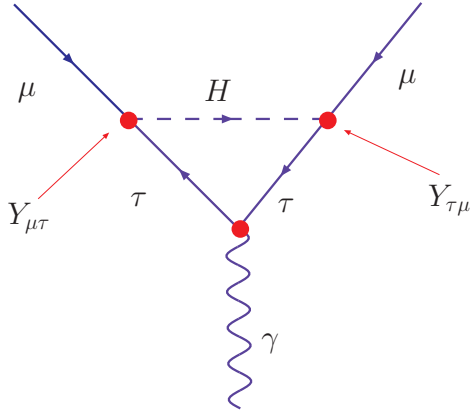}
\caption{\label{fig:anom}The contribution, $\delta a_{\mu}^{(H)}$, to the
  anomalous magnetic moment of the muon from the exchange of flavour
  changing scalars. The neutral Higgs boson can be a scalar or a pseudoscalar.}
\end{center}
\end{figure}
\end{center}

In models with more than one Higgs $SU(2)$ doublet, the exchange of
flavour changing scalars may contribute to the anomalous magnetic
moment of the muon. In the minimal $S_3$-invariant extension of the
Standard Model we are considering here, we have three Higgs
$SU(2)$ doublets, one in the singlet and the other two in the doublet
representations of the $S_3$ flavour group. The $Z_2$ symmetry
decouples the charged leptons from the Higgs boson in the $S_3$
singlet representation. Therefore, 
in the theory there are two neutral scalars and two neutral
pseudoscalars whose exchange will contribute to the anomalous magnetic
moment of the muon, in the leading order of magnitude. Since the
heavier generations have larger flavour-changing couplings, the
largest contribution comes from the heaviest charged leptons coupled
to the lightest of the neutral Higgs bosons, $\mu-\tau-H$, as shown in
Figure ~\ref{fig:anom}.
 A straightforward computation gives 
\be
\delta a_\mu^{(H)}=\frac{Y_{\mu \tau} Y_{\tau \mu}}{16
  \pi^2}\frac{m_{\mu}m_{\tau}}{M_H^2}
\left(log\left(\frac{M_{H}^2}{m_{\tau}^2}\right)-\frac{3}{2}\right).
\ee
With the help of (~\ref{y1m}) and (~\ref{y2m}) we may write $\delta
a_{\mu}^{(H)}$ as
\be
\delta a_\mu^{(H)}=\frac{m_{\tau}^2}{(246~GeV)^2}\frac{(2+\tan^2\beta)}{32
  \pi^2}\frac{m_\mu^2}{M_{H}^2}\left(log\left(\frac{M_{H}^2}{m_{\tau}^2}\right)-\frac{3}{2}\right),
\label{g-2}
\ee
in this expression, $\tan \beta=  v_{s}/v_1$, is the ratio of the
vacuum expectation values of the Higgs scalars in the singlet
representation, $v_{s}$, and
in the doublet representation, $v_{1}$, of the $S_3$ flavour group. The most
restrictive upper bound on $\tan \beta$ may be obtained from the
experimental upper bound on $Br(\mu \to 3e)$ given in (\ref{mu3e}),
and in Table~\ref{table2}, we obtain
\be
\tan \beta \leq 14.
\ee 
Substitution of this value in (\ref{g-2}) and taking for the Higgs mass
the value $M_{H}=120~GeV$ gives an estimate of the
largest possible contribution of the FCNC to the anomaly of the muon's
magnetic moment
\be
\delta a_{\mu}^{(H)}\approx 1.7 \times 10^{-10}.
\ee
This number has to be compared with the difference between the
experimental value and the Standard Model prediction for the anomaly
of the muon's magnetic moment~\cite{Jegerlehner:2007xe}
\be
\Delta a_\mu= a_\mu^{exp}-a_\mu^{SM}=(28.7 \pm 9.1 )\times 10^{-10},
\ee
which means
\be
\frac{\delta a_{\mu}^{(H)}}{\Delta a_{\mu}}\approx 0.06.
\ee
Hence, the contribution of the flavour changing neutral currents to
the anomaly of the muon's magnetic moment is smaller than or of the
order of $6\%$ of the discrepancy between the experimental value and
the Standard Model prediction. This discrepancy is of the order of
three standard deviations and quite important, but its interpretation
is compromised by uncertainties in the computation of higher order
hadronic effects arising mainly from three-loop vacuum polarization
effects, $a_{\mu}^{VP}(3,had)\approx-1.82\times 10^{-9}$~\cite{Erler:2006vu}, and
from three-loop contributions of hadronic light by light type,
$a_{\mu}^{LBL}(3,had)\approx 1.59\times 10^{-9}$~\cite{Erler:2006vu}. As explained above, the
contribution to the anomaly from flavour changing neutral currents in
the minimal $S_{3}$-invariant extension of the Standard Model,
computed in this work is, at most, $6\%$ of the discrepancy between
the experimental value and the Standard Model prediction for the anomaly, and is of the same order of magnitude as
the uncertainties in the higher order hadronic contributions, but
still, it is not negligible and is certainly compatible with the best,
state of the art, experimental measurements and theoretical computations.

\section{Conclusions}
In the minimal $S_{3}$-invariant extension of the SM the flavour symmetry group $Z_{2}
\times S_{3}$ relates the mass spectrum and mixings. This allowed us
to compute the neutrino mixing matrix explicitly in terms of the
masses of the charged leptons and neutrinos~\cite{Felix:2006pn}. In this model, the magnitudes
of the three mixing angles are determined by the interplay of the
flavour $S_{3}\times Z_{2}$ symmetry, the see-saw mechanism and the
lepton mass hierarchy. We also found that $V_{PMNS}$ has three CP
violating phases, one Dirac phase
$\delta=\delta_{\nu}-\delta_{e}$ and two 
Majorana phases, $\alpha$ and
$\beta$, that are functions of the neutrino masses, and another phase
$\phi_{\nu}$ which is independent of the Dirac phase. The numerical
values of the reactor, $\theta_{13}$, and the atmospheric,
$\theta_{23}$, mixing angles are determined by the masses of the
charged leptons only, in very good agreement with the experiment. The
solar mixing angle $\theta_{12}$ is almost insensitive to the values
of the masses of the charged leptons, but its experimental value
allowed us to fix the scale and origin of the neutrino mass spectrum,
which has an inverted hierarchy, with the values $|m_{\nu_{2}}|=0.056eV$,
$|m_{\nu_{1}}|=0.055eV$ and $|m_{\nu_{3}}|=0.022eV$.
We also obtained explicit expressions for the matrices of the Yukawa
couplings of the lepton sector parametrized in terms of the charged lepton masses and the
VEV's of the neutral Higgs bosons in the $S_3$-doublet
representation. These Yukawa matrices are closely related to the
fermion mass matrices and have a structure of small and very small
entries reflecting the observed charged lepton mass hierarchy. With
the help of the Yukawa matrices, we computed the branching ratios of a
number of FCNC processes and found that the branching ratios of all
FCNC processes considered are strongly suppressed by powers of the
small mass ratios $m_e/m_\tau$  and $m_\mu/m_\tau$, and by the ratio
$\left(m_\tau/M_{H_{1,2}}\right)^4$, where $M_{H_{1,2}}$ is the mass
of the neutral Higgs bosons in the $S_3$-doublet. Taking for
$M_{H_{1,2}}$  a very conservative value ($M_{H_{1,2}}\approx
120~GeV$), we found that the numerical values of the branching ratios
of the FCNC in the leptonic sector are well below the corresponding
experimental upper bounds by many orders of magnitude. It has already been argued that small FCNC
processes mediating non-standard quark-neutrino interactions could be
important in the theoretical description of the gravitational core
collapse and shock generation in the explosion stage of a
supernova~\cite{Raffelt:2007nv,Amanik:2004vm,valle-nu}.
Finally, the contribution of the flavour changing neutral currents to
the anomalous magnetic moment of the muon is small but non-negligible
and it is compatible with the best, state of the art measurements and
theoretical computations.
\section{Acknowledgements}
We thank Prof. Jens Erler and Dr. Genaro Toledo-Sanchez for
helpful discussions about $g-2$. This work was partially supported by CONACYT M\'exico under contract
No 51554-F and by DGAPA-UNAM under contract PAPIIT-IN115207-2.

\end{document}